\documentclass[11pt]{article}

\usepackage{fullpage}

\usepackage{mathrsfs}
\usepackage{latexsym}
\usepackage{amsmath}
\usepackage{amssymb}
\usepackage{amsfonts}
\usepackage{ifthen}

\def\Natural{\mathbb{N}}

\def\01{\{0,1\}}
\newcommand{\ceil}[1]{\lceil{#1}\rceil}

\newcommand{\ket}[1]{|#1\rangle}

\newcommand{\Prob}{\mbox{\rm Prob}}

\newcounter{protoCount}
\newcounter{protoList}
\newsavebox{\tmpbox}
\newlength{\protobox}
\newenvironment{protocol}[2]{
\bigskip
\addtocounter{protoCount}{1}
\noindent \begin{lrbox}{\tmpbox}
\setlength{\protobox}{\textwidth}
\addtolength{\protobox}{-0.5cm}
\begin{minipage}[c]{\protobox}
\begin{bfseries}Protocol \theprotoCount: #1\end{bfseries}
\ifthenelse{\equal{#2}{\empty}}{}{\\Prerequisite: #2}
\begin{list}{\begin{bfseries}\arabic{protoList}:\end{bfseries}}
{\usecounter{protoList}}
}{
\end{list}
\end{minipage}\end{lrbox}
\fbox{\usebox{\tmpbox}}
\bigskip
}

\newtheorem{definition}{Definition}
\newtheorem{theorem}{Theorem}
\newtheorem{lemma}{Lemma}

\newtheorem{corollary}{Corollary}

\newenvironment{proof}
{\noindent {\bf Proof. }}
{{\hfill $\Box$}\\
 \smallskip}

\title{Quantum Anonymous Transmissions}
\author{Matthias Christandl\thanks{Supported by the EU project RESQ IST-2001-37559, a DAAD
Doktorandenstipendium and the U.K.~Engineering and Physical Sciences
Research Council.}\\
Centre for Quantum Computation\\
University of Cambridge\\
matthias.christandl@qubit.org
\and
Stephanie Wehner\thanks{Supported
by the EU project RESQ IST-2001-37559 and the NWO vici project 2004-2009. Part of
this work was done while visiting CQC Cambridge.}\\
CWI, Amsterdam\\
wehner@cwi.nl
}

\begin{document}
\maketitle
\begin{abstract}
We consider the problem of hiding sender and receiver of
classical and quantum bits (qubits), even if all physical transmissions can
be monitored. We present a quantum protocol for sending and
receiving classical bits anonymously, which is completely
traceless: it successfully prevents later reconstruction of the
sender. We show that this is not possible classically. It appears
that entangled quantum states are uniquely suited for traceless
anonymous transmissions. We then extend this protocol to send and
receive qubits anonymously. In the process we introduce a new
primitive called anonymous entanglement, which may be useful in
other contexts as well.
\end{abstract}

\section{Introduction}

In most cryptographic applications, we are interested in ensuring
the secrecy of data. Sender and receiver know each other, but are
trying to protect their data exchange from prying eyes.
Anonymity, however, is the secrecy of identity. Primitives to hide
the sender and receiver of a transmission have received
considerable attention in classical computing. Such primitives
allow any member of a group to send and receive data
anonymously, even if all transmissions can be monitored.
They play
an important role in protocols for electronic
auctions~\cite{stajano99cocaine}, voting protocols and sending
anonymous email~\cite{chaum:mixnet}. Other applications allow
users to access the Internet without revealing their own
identity~\cite{reiter:crowds},~\cite{pipenet} or, in combination
with private information retrieval, provide anonymous
publishing~\cite{dingledine:thesis}. Finally, an anonymous
channel which is completely immune to any active
attacks, would be a powerful primitive. It has been shown how two
parties can use such a channel to perform
key-exchange~\cite{alpern:keyless}.

\subsection{Previous Work}

A considerable number of classical schemes have been suggested for
anonymous transmissions. An unconditionally secure classical
protocol was introduced by Chaum~\cite{chaum:dc} in the context of
the Dining Cryptographers Problem. Since this protocol served as
an inspiration for this paper, we briefly review it here. A group
of cryptographers is assembled in their favorite restaurant. They
have already made arrangements with the waiter to pay anonymously,
however they are rather anxious to learn whether one of them is
paying the bill, or whether perhaps an outside party such as the
NSA acts as their benefactor. To resolve this question, they all
secretly flip a coin with each of their neighbours behind the menu
and add the outcomes modulo two. If one of them paid, he inverts
the outcome of the sum. They all loudly announce the result of
their computation at the table. All players can now compute the
total sum of all announcements which equals zero if and only if
the NSA pays. This protocol thus allows anonymous transmission of
one bit indicating payment. A network based on this protocol is
also referred to as a DC-net\index{DC-net}. Small scale practical
implementations of this protocol are known~\cite{martin:thesis}.
Boykin~\cite{boykin:thesis}\index{Boykin, P. O.} considered a
quantum protocol to send classical information anonymously where
the players distribute and test pairwise shared EPR pairs, which
they then use to obtain key bits. His protocol is secure in the
presence of noise or attacks on the quantum channel. Other
anonymity related work was done by M\"uller-Quade and
Imai~\cite{AOT} in the form of anonymous oblivious transfer.

In practice, two other approaches are used, which do not aim for
unconditional security: First, there are protocols which employ a
trusted third party. This takes the form of a trusted proxy server
\cite{anonymizer},~\cite{penetfi}, forwarding messages while
masking the identity of the original sender. Secondly, there are
computationally secure protocols using a chain of forwarding
servers. Most notably, these are protocols based on so-called
mixing techniques introduced by
Chaum~\cite{chaum:mixnet}\index{MixNet}\index{Chaum, D.}, such as
Webmixes~\cite{webmixes} and ISDN-Mixes~\cite{pfitzmann:isdn}.
Here messages are passed through a number of proxies
which reorder the messages; hence the name MixNet. The goal of
this reordering is to ensure an observer cannot match in- and
outgoing messages and thus cannot track specific messages on their
way through the network. Public Key Encryption is then used
between the user and the different forwarding servers to hide the
contents of a message. Several implemented systems, such as
Mixmaster~\cite{mixmaster},
 PipeNet~\cite{pipenet}, Onion Routing~\cite{onion} and Tor~\cite{tor:design,tor} employ layered encryption:
the user successively encrypts the message with the public keys of
all forwarding servers in the chain. Each server then ``peels
off'' one layer, by decrypting the received data with its own
secret key, to determine the next hop to pass the message to. The
Crowds~\cite{reiter:crowds} system takes another approach. Here
each player acts as a forwarding server himself. He either sends
the message directly to the destination, or passes it on to
another forwarding server with a certain probability. The aim is
to make any sender within the group appear equally probable for an
observer. Various other protocols using forwarding
techniques are known. Since our focus lies on unconditionally
secure protocols, we restrict ourselves to this brief
introduction. More information can be found in the papers by
Goldberg and Wagner~\cite{goldberg:privacy},~\cite{goldberg:privacynew}\index{Goldberg, I.}\index{Wagner, D.} and
in the PhD thesis of Martin
\cite[Chapter 2 and 3]{martin:thesis}.\index{Martin, D.}

Note that a DC-net computes the parity of the players inputs.
Sending classical information anonymously can thus be achieved
using secure multi-party computation which has received
considerable attention
classically~\cite{goldreich:mental},~\cite{chaum:multiparty}.
 Quantum secure multi-party computation
has been considered for the case that the players hold quantum inputs and
each player receives part of the output~\cite{crepeau:multiparty}.
Our protocol for sending qubits anonymously does not form
an instance of general quantum secure multi-party computation, as
we only require the receiver to obtain the qubit sent. Other
players do not share part of this state. Instead, the receiver
of the state should remain hidden.

\subsection{Contribution}

Here we introduce quantum protocols to send and receive classical
and quantum bits anonymously. We first consider a protocol that
allows $n$ players to send and receive one bit of classical
information anonymously using one shared entangled state
$\ket{\Psi} = (\ket{0}^{\otimes n} + \ket{1}^{\otimes n}
)/\sqrt{2}$ and $n$ uses of a broadcast channel\index{broadcast}.
Given these resources, the protocol is secure against collusions
of up to $n-2$ players: the collaborators cannot learn anything
more by working together and pooling their resources. 

The most notable property of our protocol for anonymous
transmissions of classical data is that it is traceless as defined
in Section~\ref{traceless}. This is related to the notion of
incoercibility in secure-multi party
protocols~\cite{canetti:coerce}.  Informally, a protocol is
incoercible, if a player cannot be forced to reveal his true input
at the end of the protocol. When forced to give up his input,
output and randomness used during the course of the protocol, a
player is able to generate fake input and randomness instead, that
is consistent with the public transcript of communication. He can
thus always deny his original input. This is of particular
interest in secret voting to prevent vote-buying. Other examples
include computation in the presence of an authority, such as the
mafia, an employer or the government, that may turn coercive at a
later point in time. In our case, incoercibility means that a
player can always deny having sent. A protocol that is traceless,
is also incoercible. However, a traceless protocol does not even
require the player to generate any fake randomness. A sender can
freely supply a fake input along with the true randomness used
during the protocol without giving away his identity, i.e. his role
as a sender during the protocol. This can be
of interest in the case that the sender has no control over which
randomness to give away. Imagine for example a burglar sneaking in
at night to obtain a hard disk containing all randomness or the
sudden seizure of a voting machine. As we show, the property
traceless of our protocol contrasts with all classical protocols
and provides another example of a property that cannot be achieved
classically. The protocols suggested in \cite{boykin:thesis} are
not traceless, can, however, be modified to exhibit this property.

Clearly, in 2005 the group of dinner guests is no longer content
to send only classical bits, but would also like to send qubits
anonymously. We first use our protocol to allow two anonymous
parties to establish a shared EPR pair\index{EPR pair}. Finally,
we use this form of anonymous entanglement to hide the sender and
receiver of an arbitrary qubit. These protocols use the same
resource of shared entangled states $\ket{\Psi}$ and a broadcast
channel. 

\subsection{Outline}
Section~\ref{prelim} states the resources used in the protocol,
necessary definitions and a description of the model. In
Section~\ref{sec-class-limit} we derive limitations on classical
protocols. Section~\ref{classical} then presents a quantum
protocol for sending classical bits anonymously.
Section~\ref{quantum} deals with the case of sending qubits
anonymously and defines the notion of anonymous entangle\-ment.
Multiple simultaneous senders are considered in
Section~\ref{multiple senders}. 

\section{Preliminaries}\label{prelim}

\subsection{Definitions and Model}\label{traceless}

We will consider protocols among a set of $n$ players who are
consecutively numbered. The players may assume a distinct role in
a particular run of the protocol. In particular, some players
might be \emph{senders} and others \emph{receivers} of \emph{data
items}. In our case, a data item $d$ will be a single bit or a qubit.
We use the verb \emph{send} to denote transmission of a
data item via the anonymous channel and \emph{transmit} to
denote transmission of a \emph{message} (here classical bits) via
the underlying classical message passing network\footnote{A
network of pairwise communication channels between the
players.} or via the broadcast channel given in Definition
\ref{def-FGMR}.

Anonymity is the secrecy of identity. Looking at data
transmissions in particular, this means that a sender stays
\emph{anonymous}, if no one can determine his identity within the
set of possible senders. In particular, the receiver himself should
not learn the sender's identity either. Likewise, we define anonymity for the
receiver. In all cases that we consider below, the possible set of
senders coincides with the possible set of receivers. 
The goal of an \emph{adversary} is to determine the
identity of the sender and/or receiver. To this end he can choose
to \emph{corrupt} one or more players: this means he can take complete control
over such players and their actions. Here, we only consider a non-adaptive
adversary, who chooses the set of players to corrupt before the start
of the protocol. In addition, the adversary is allowed to monitor all
physical transmissions: he can follow the path of all messages, reading
them as desired. Contrary to established literature, we here give the
adversary one extra ability: After completion of the protocol, the adversary 
may \emph{hijack} any number of players. This means that he can break 
into the system of a hijacked player and learn all randomness this player used during the protocol.
However, he does not learn the data item $d$ or the role this player played
during the protocol. 
In a DC-net, for example, the randomness
are the coin flips performed between two players.
The adversary may then try to use this additional information
to determine the identity of the sender and/or receiver. 
We return to the concept of hijacking in Section~\ref{traceless}. 
In this paper,
we are only interested in unconditional security and thus consider an
unbounded adversary.
We call a player \emph{malicious} if he is corrupted 
by the adversary. A malicious player may deviate from the protocol
by sending alternate messages. We call a player \emph{honest}, if
he is not corrupted and follows the protocol. If $t > 1$ players
are corrupted, we also speak of a \emph{collusion} of $t$ players.

Let $V$ denote the set of all players. Without loss of generality,
a \emph{protocol} is a sequence of $k$ rounds,
where in each round the players, one after another, transmit
one message.
We use $c_{jm}$ to denote the message transmitted by player $m$ in round $j$.
The total communication during the protocol is thus given by the sequence
$C = \{c_{jm}\}_{j=1, m=1}^{k, n}$ of $nk$ messages. Note that we do not
indicate the receiver of the messages. 
At the beginning of the protocol, the players may have access to private
randomness and shared randomness among all players, or a subset of players.
In addition, each player may generate local private randomness during the course of
the protocol. We use $g_{jm}$ to denote the random string held by player
$m$ in round $j$. A player cannot later delete $g_{jm}$.
Let $G_m = \{g_{jm}\}_{j=1}^k$ be the combined
randomness held by player $m$. Similarly, we use $G = \{G_m\}_{m=1}^n$
to denote the combined randomness held by all players. 
Note that the data item $d$ player $m$ wants to send and his role in the protocol (sender/receiver/none)
are excluded from $G_m$. In the following definitions, we exclude the trivial 
case where the sender or receiver are known beforehand, and where the sender
is simultaneously the receiver.

It is intuitive that a protocol preserves the anonymity of a
sender, if the communication does not change the a priori
uncertainty about the identity of the sender. Formally:

\begin{definition}\index{anonymous}
A $k$-round protocol $P$ allows a sender $s$ to be
\emph{anonymous}, if for the adversary who corrupts $t \leq n-2$ players
$$\max_{S} \Prob[S=s|G^t,C]=\max_{S}
\Prob[S=s]=\frac{1}{n-t}$$ where the first maximum is taken over
all random variables $S$ which depend only on the sequence of all messages, $C$, and on the set of randomness held by the corrupted players,
$G^t = \{G_m\}_{m \in E}$. Here, $E \subset V \backslash \{s\}$ is the set of players corrupted by the 
adversary; to exclude the trivial case where the sender $s$ himself is corrupted by the adversary.
A protocol $P$ that allows a sender to be anonymous achieves \emph{sender
anonymity}.
\end{definition}
Similarly, we define the anonymity of a receiver:

\begin{definition}
A $k$-round protocol $P$ allows a receiver $r$ to be
\emph{anonymous}, if for the adversary who corrupts $t \leq n-2$ players
$$\max_{R} \Prob[R=r|G^t,C]=\max_{R}
\Prob[R=r]=\frac{1}{n-t}$$ where the first maximum is taken over
all random variables $R$ which depend only on the sequence of all messages, $C$, and on the set of randomness held by the corrupted players,
$G^t = \{G_m\}_{m \in E}$. Here, $E \subset V \backslash \{r\}$ is the set of players corrupted by the
adversary; to exclude the trivial case where the receiver $r$ himself is corrupted by the adversary.
A protocol
$P$ that permits a receiver to be anonymous achieves
\emph{receiver anonymity}.
\end{definition}
Note that protocols to hide the sender
and receiver may not protect the data item sent. In particular
there \emph{could} be more players receiving the data item, even
though there is only one receiver, which is determined before the
protocol starts. The definition implies that the data sent via the
protocol does not carry any compromising information itself.

All known protocols for sender and receiver anonymity achieving
information theoretic security need a reliable broadcast channel
~\cite{maurer:byzantine-eurocrypt}. We will also make use of this
primitive:

\begin{definition}[FGMR~\cite{maurer:byzantine-eurocrypt}] \label{def-FGMR}
A protocol among $n$ players such that one distinct player s (the
sender) holds an input value $x_s \in L$ (for some finite domain
$L$) and all players eventually decide on an output value in $L$
is said to achieve \emph{broadcast} (or \emph{Byzantine
Agreement}) if the protocol guarantees that all honest players
decide on the same output value $y \in L$, and that $y = x_s$
whenever the sender is honest.
\end{definition}

Informally, we say that a protocol is traceless, if it remains
secure even if we make all resources available to an adversary at
the end of the protocol. Consider for example the DC-net protocol
discussed earlier. Imagine a curious burglar sneaking into the
restaurant at night to gather all coin flips our group of
cryptographers performed earlier on from the tapes of the security
cameras. A protocol is traceless, if it can withstand this form of
attack. 

We model this type of attack by granting the adversary one additional 
ability.
After completion of the protocol, we allow the adversary to hijack
any number of players. If an adversary \emph{hijacks} player $m$, he breaks
into the system and learns all randomness $G_m$ used by this player.
In this paper, we allow the adversary to hijack all players after
completion of the protocol. The adversary then learns all randomness
used by the players, $G$. Nevertheless, we want him to remain ignorant
about the identity of the sender and receiver. Formally,

\begin{definition}
A $k$-round protocol $P$ with sender $s$ which achieves sender
anon-ymity is \emph{sender traceless}, if for the adversary who corrupts 
any $t \leq n-2$ 
players and, after completion
of the protocol, hijacks all players
$$\max_{S} \Prob[S=s|G,C]=\max_{S}
\Prob[S=s]=\frac{1}{n-t}$$ where the first maximum is taken over all
random variables $S$ which depend only on the sequence of all messages, $C$, and on the set of randomness held by all players,
$G$.
\end{definition}

Likewise, change of sender $s$ with receiver $r$, we
define the property traceless for receiver anonymous protocols.
Recall that $G$ and $C$ do not contain the data item $d$ that was sent
or the roles the players assumed during the course of the protocol.

\subsection{Limitations on Traceless Protocols}
\label{sec-class-limit}

Intuitively, we cannot hope to construct a classical protocol
which is traceless and at the same time allows the receiver to
learn what was sent: The only way data $d$ can be send classically
is by transmitting messages over the underlying network. If, however,
an adversary has all information except the player's input and all
communication is public, he can simply check the
messages transmitted by each player to see if they ``contain''
$d$. 

\begin{theorem}
Let $P$ be a classical protocol with one sender and one receiver
such that for all data items $d \in D$ with $|D| \geq 2$ the
following holds: the sender of $d$ stays anonymous and the
receiver knows $d$ at the end of the protocol. Then $P$ is not
sender traceless.
\end{theorem}

\begin{proof}
Let us assume by contradiction that the protocol is traceless.
Without loss of generality, a player who is not the sender has
input $d_0 \in D$ to the protocol. Let $d \in D$ be the data item
that the sender $s$ wants to send. We assume that all but one players are
honest during the run of the protocol. We would like to emphasize
that the only information that is not written down, is in fact the
data item $d$ of the sender.

The adversary corrupts one player.
After completion of the protocol, he hijacks all players. He thus
has access to all randomness and communication. Since a
traceless protocol must resist the corruption of any player, it
must also resist the corruption of the receiver. We therefore
assume for the remainder of the proof that the adversary corrupts
the receiver.

Let us consider step $j$ in the protocol, where player $m$ has
total information $g_{jm}$ and sends communication $c_{jm}$. Note
that $c_{jm}$ may only depend on the previous communication,
$g_{jm}$, $j$, the number $m$ and the role of the player $m$, i.e.
whether $m$ is sender, receiver or neither of them. If $m=s$, then
the communication may also depend on $d$. Since the adversary has
corrupted the receiver, and since there is only one receiver, the
adversary knows that $m$ is either a normal player or the sender.
Note that since the adversary corrupted the receiver, he also knows 
the value of $d$.

After the protocol, the adversary, having access to $G$ and $C$,
can now calculate the messages that player $m$ should have sent in
round $j$ depending on whether 
\begin{enumerate}
\item $m$ was not sender or receiver, or,
\item $m$ was the sender and sent item $d$.
\end{enumerate}
The messages are calculated as follows: In case 1, the adversary 
simulates the actions of player $m$ as if $m$ was neither sender
nor receiver. This is possible, since the adversary has access
to all randomness and all communication. In case 2, the adversary
simulates the actions of $m$ as if $m$ was the sender and sent
data item $d$. Let $\{f^1_{jm}\}_j$,$\{f^2_{jm}\}_j$ denote the set of messages
resulting from the simulations of cases 1 and 2 respectively.
The adversary now checks whether the set of observed messages
$\{c_{jm}\}_j = \{f^1_{jm}\}_j$ or $\{c_{jm}\}_j = \{f^2_{jm}\}_j$.
If the first equality holds he concludes that 
$s \neq m$, and for the second that $s = m$.

By assumption, the protocol is traceless for all $d$. Thus, the
message computed for case 2) must be identical to the
message computed for case 1) for all $d$, since otherwise the adversary could
determine the sender $s$. This must hold for all steps $j$. But in
this case the strategy the sender follows must be the same for
both $d=d_0$ and $d \neq d_0$. Hence it cannot have been possible
for $r$ to have obtained the value of $d$ in the first place and
we have a contradiction to the assumption that the protocol 
achieves a transfer for all elements of a set $D$ with $|D| \geq
2$.
\end{proof}
\noindent
Note that we make the assumption that there is exactly one
receiver which is determined before the start of the protocol.
Other players might still obtain the data item, as this is not a
statement about the security of the message but merely about
anonymity.

\subsection{Limitations on Shared Randomness}

In this section, we take a look at how many privately shared
random bits are needed in order to perform anonymous
transmissions. We thereby only consider unconditionally secure
classical protocols based on privately shared random bits, such as
for example the DC-net. In the following, we will view the players
as nodes in an undirected graph. The notions of ``nodes in a
key-sharing graph'' and ``players'' are used interchangeably.
Similarly, edges, keys and private shared random bits are the
same. Again, regard the broadcast channel as an abstract
resource.
\begin{definition}\index{key-sharing graph} The
undirected graph $G = (V,E)$ is called the \emph{key-sharing
graph} if each node in $V$ represents exactly one of the players
and there is an edge between two nodes $i$ and $j$ if and only if
$i$ and $j$ share one bit of key $r_{i,j}$.
\end{definition}

We first note that for any protocol $P$ that achieves sender
anonymity, where the only resource used by the $n$ participating
players is pairwise shared keys, a broadcast channel and public
communication, the form of the key-sharing graph $G = (V,E)$ is
important: 

\begin{lemma}\label{partition}
In any protocol $P$ to achieve sender anonymity among $n$ players,
where the only resource available to the players is pairwise
shared keys, a broadcast channel and public communication, a
collusion of $t$ players can break the sender's anonymity, if the
corresponding collection of $t$ nodes partitions the key-sharing
graph $G = (V,E)$.
\end{lemma}
\begin{proof}
$t$ colluding nodes divide the key-sharing graph into $s$ disjoint
sets of nodes $\{S_1,\ldots,S_s\}$. Note that there is no edge
connecting any of these sets, thus these sets do not share any
keys. Now suppose that sender anonymity is still
possible. Let $k_i \in S_i$ and $k_j \in S_j$ with $i \neq j$ be
two nodes in different parts of the graph. Using a protocol achieving sender anonymity
it is now possible to establish a secret bit between
$k_i$ and $k_j$~\cite{alpern:keyless}:
Nodes $i$ and $j$ each generate $n$ random bits: $r_i^1,\ldots,r_i^n$ and
$r_j^1,\ldots,r_j^n$. Node $i$ now announces $n$ data of
the form: ``Bit $b_k$ is $r_i^k$'' for $1 \leq k \leq n$
using the protocol for sender anonymity. Likewise, node $j$
announces ``Bit $b_k$ is $r_j^k$'' for $1 \leq k \leq n$.
Nodes $i$ and $j$ now discard all bits for which $r_i^k = r_j^k$ and use the
remaining bits as a key.  Note that
an adversary can only learn whether $b_k = r_i^k$ or $b_k = r_j^k$ if the two
announcements are the same. If
$r_i^k \neq r_j^k$, the adversary does not learn who has which bit.

However, there is no channel between $S_i$ and $S_j$
that is not monitored by the colluding players. Thus, it cannot be
possible to establish a secret bit between $k_i$ and $k_j$, since
the only communication allowed is classical and
public~\cite{nielsen&chuang:qc}. This establishes the contradiction and shows
that the sender's anonymity can be broken if the graph can be
partitioned.
\end{proof}
\noindent
Furthermore, note that each player $j$ needs to share one bit of key with at least
two other players. Otherwise, his anonymity can be compromised. We can phrase this in terms
of the key-sharing graph as

\begin{corollary}\label{degree}
Each node $j \in V$ of the key-sharing graph $G = (V,E)$, used by
a protocol $P$ for anonymous transmissions, where the only
resource available to the $n$ players is pairwise shared keys, a
broadcast channel and public communication, must have degree $d
\geq 2$.
\end{corollary}
\begin{proof}
Suppose on the contrary, that an arbitrary node $j$ has degree 1: it has only one outgoing
edge to another node $k$. Clearly, node $k$ can partition the key-sharing graph into two disjoint
sets $S_1 = \{j\}$ and $S_2 = V\setminus \{j,k\}$. By Lemma~\ref{partition}, node $k$ can break
$j$'s anonymity.
\end{proof}

\begin{corollary}
Any protocol $P$ that achieves sender anonymity, where no players
collude and the only resource available to the $n$ players is
pairwise shared keys, a broadcast channel and public communication,
needs at least $n$ bits of pairwise shared keys.
\end{corollary}
\begin{proof}
Consider again the key-sharing graph $G = (V,E)$.
Suppose on the contrary, that only $k < n$ bits of shared
keys are used. Then there must be at least one node of degree $1$ in the graph.
Thus, by Corollary~\ref{degree} at most $n$ bits of shared keys are necessary.
\end{proof}

\begin{corollary}
Any protocol $P$ that achieves sender anonymity and is resistant
against collusions of $t < n - 1$ players, where the only
resources available to the $n$ players are pairwise shared keys, a
broadcast channel and public communication, needs at least
$n(n-1)/2$ bits of pairwise shared keys.
\end{corollary}
\begin{proof}
Again consider the key-sharing graph $G$. Suppose on the contrary, that only $k < n(n-1)/2$ bits
of shared keys are used. However, then there are only $k < n(n-1)/2$ edges in a graph of $n$ nodes.
Then $G$ is not fully connected and there is a set of $t = n-2$ colluding nodes which can partition the key-sharing graph.
By Lemma~\ref{partition}, they can then break the sender's anonymity. Thus $n(n-1)/2$ bits of pairwise
shared key are necessary to tolerate
up to $t < n-1$ colluding players.
\end{proof}

\subsection{Quantum Resources}
We assume familiarity with the quantum
model~\cite{nielsen&chuang:qc}. The fundamental resource used in
our protocols are $n$-party shared entangled
states\index{state!entangled} \index{entangled states} of the form
$$
\ket{\Psi} = 
\frac{1}{\sqrt{2}}(\ket{0^n} + \ket{1^n})
\equiv 
\frac{1}{\sqrt{2}}( \ket{0}^{\otimes n} + \ket{1}^{\otimes n}). 
$$
These are commonly known as generalized GHZ states~\cite{GHZ}. By
``shared''\index{shared} we mean that each of the $n$ players
holds exactly one qubit of $\ket{\Psi}$. They could have obtained
these states at an earlier meeting or distribute and test them
later on.

The key observation used in our protocols is the fact that phase flips\index{phase flip} and rotations\index{rotation} 
applied by the individual players have the same effect on the global state no matter
who applied them. Consider for example the phase flip defined by
$$
\sigma_z = \left(\begin{array}{cc}1 & 0\\0 & -1\end{array}\right). 
$$  
If player number $i$ applies this transformation\index{transformation} to his state, the global transformation
is
$U_i = I^{\otimes (i-1)} \otimes \sigma_z \otimes I^{\otimes (n - i)}$,
where $I$ is the identity transform.
We now have $\forall i \in \{1,\ldots,n\}: U_i\ket{\Psi} = (\ket{0^n} - \ket{1^n})/\sqrt{2}$. Note that this
transformation takes place ``instantaneously'' and no communication is necessary.

\section{Traceless Quantum Protocols}

\subsection{Model}\label{model}

To obtain traceless anonymous transmissions we 
allow the players to have access to a generalized GHZ state.
We assume that the $n$ players have access to the following
resources:
\begin{enumerate}
\item $n$-qubit shared entangled states $\ket{\Psi} = (\ket{0^n} + \ket{1^n})/\sqrt{2}$ on
which the players can perform arbitrary measurements.
\item A reliable broadcast channel.
\end{enumerate}

\subsection{Sending Classical Bits}\label{classical}

To start with, we present a protocol to send a classical bit
$b$ anonymously, if the $n$ players share an $n$-qubit
entangled state $\ket{\Psi}$. For now, we assume that only one
person wants to send in each round of the protocol and deal
with the case of multiple senders later on.
We require our protocol to have the following properties:
\begin{enumerate}
\item (Correctness) If all players are honest, they receive the
data item $d$ that was sent by the sender. If some players are malicious, 
the protocol aborts or all honest players receive the
same data item $\tilde{d}$, not necessarily equal to $d$.
\item (Anonymity) If up to $t \leq n-2$ players are malicious, the sender and 
receiver stay anonymous. 
\item (Tracelessness) The protocol is sender and receiver traceless.
\end{enumerate}

\subsubsection{Protocol}

Let's return to the original dinner table scenario described
earlier. Suppose Alice, one of the dinner guests, wishes to send a
bit $d \in D= \01$ anonymously. For this she uses the following
protocol:

\begin{protocol}{ANON($d$)}{Shared state $(\ket{0^{n}} + \ket{1^{n}})/\sqrt{2}$}\index{ANON protocol}
\item Alice applies a phase flip $\sigma_z$ to her part of the state if $d = 1$ and does
nothing otherwise.
\item Each player (incl. Alice):
\begin{list}{-}{}
\item Applies a Hadamard transform to his/her qubit.
\item Measures his/her qubit in the computational basis.
\item Broadcasts his/her  measurement result.
\item Counts the total number of 1's, $k$, in the $n$ measurement outcomes.
\item If $k$ is even, he/she concludes $d = 0$, otherwise $d = 1$.
\end{list}
\item The protocol aborts if one of more players do not use the broadcast
channel.
\end{protocol}

\subsubsection{Correctness}

First of all, suppose all parties are honest. Since Alice applies
the phase flip $\sigma_z$ depending on the value of the bit $d$
she wishes to send, the players obtain the state $(\ket{0^n} +
\ket{1^n})/\sqrt{2}$ if $d = 0$ and $(\ket{0^n} -
\ket{1^n})/\sqrt{2}$ if $d = 1$. By tracing out the other players'
part of the state, we can see that no player can determine on his
own whether the phase of the global state has changed. We
therefore require the players to first apply a
Hadamard\index{Hadamard transform} transform $H$ to their qubit.
This changes the global state such that we get a superposition of
all strings $x \in \01^n$ with an even number of 1's for no phase
flip and an odd number of 1's if a phase flip has been applied:
\begin{eqnarray*}
&&H^{\otimes n}\left(\frac{1}{\sqrt{2}}(\ket{0^n} + (-1)^d
\ket{1^n})\right) =\\
&=& \frac{1}{\sqrt{2^{n+1}}}\left(\sum_{x \in \01^n} \ket{x} + (-1)^d \sum_{x \in \01^n} (-1)^{|x|} \ket{x}\right)\\
&=& \frac{1}{\sqrt{2^{n+1}}}\sum_{x \in \01^n} (1 + (-1)^{d \oplus
|x|}) \ket{x},
\end{eqnarray*}
where $|x|$ denotes the Hamming weight of the string $x$. Thus we
expect an even number of 1's if $d = 0$ and an odd number of 1's
if $d = 1$. The players now measure\index{measurement} their part
of the state and announce the outcome. This allows each player to
compute the number of 1's in the global outcome, and thus $d$. If
more than one player had applied a phase flip, ANON computes the
parity of the players inputs. Broadcasting all measurement results
needs $n$ uses of a broadcast channel.

Now suppose that some of the players are malicious. Recall that we
assume that the players use a reliable broadcast channel. This
ensures an honest player obtains the same value for the
announcement. Thus two honest parties will never compute a
different value for the sent data item $d$. Further, note that it may always
be possible that one or more malicious players do not use the
broadcast channel. This consequently results in an abort of the
protocol. We conclude that the correctness condition is satisfied.

\subsubsection{Anonymity}\label{anon security}

As we noticed in Section~\ref{prelim}, the resulting global state
is independent of the identity of the person applying the phase
flip. Since a phase flip is applied locally, no transmissions are
necessary to change the global state. Subsequent transmissions are
only dependent on the global state. Since this global state is
invariant under an arbitrary permutation of the honest players and
since the communication of the individual players depends only on
their part of the states, the total communication during a run of
the protocol $P$ where player $m$ sends $d$, is independent of the
role of the player. If the sender is not one of the colluding
players, then for the set of colluding players, all other players
are equally likely to be sender. This is precisely the definition
of sender anonymity.
A receiver may be specified. His anonymity is then given directly
as every player obtains the bit sent.

Note that a player deviating from the protocol by inverting his
measurement outcome or applying a phase flip himself will only
alter the outcome, but not learn the identity of the sender.
The same discussion holds when the protocols
is executed multiple times in succession or parallel.

\subsubsection{Tracelessness}

The most interesting property of our quantum protocol is that it
is completely traceless: The classical communication during the
protocol is solely dependent on the global state, which is the
same no matter who the sender is. This means that Alice'
communication is independent of her bit $d$. The randomness is
now determined by the measurement results of the global state,
which has already been altered according to the players inputs.
Thus, the traceless condition is satisfied, because there is thus 
no record of Alice sending.

We believe that the tracelessness is a very intuitive
property of the quantum state, as sending $d$ simply changes the
overall probability distribution of measurement outcomes instead
of the individual messages of the sender. Note, however, that if
we had first measured the state $\ket{\Psi}$ in the Hadamard basis
to obtain classical information and then allowed the sender to
invert the measured bit to send $d=1$, our protocol would no
longer be traceless. We leave no record of Alice' activity in
the form of classical information. Alice can later always deny
that she performed the phase flip. Whereas this is stronger than
classical protocols, it also makes our protocol more prone to
disruptors. Unlike in the classical scenario, we cannot employ
mechanisms such as traps suggested by Chaum~\cite{chaum:dc}, and
Waidner and Pfitzmann~\cite{dcdisco}, to trace back disruptors. If
one of our players is determined to disrupt the channel by, for
example, always applying a phase flip himself, we are not able to
find and exclude him from the network.

\subsection{Anonymous Entanglement}\index{anonymous!entanglement}\index{entanglement!anonymous}
The dinner guests realize that if they could create entanglement with any of the other players anonymously,
 they could teleport a quantum state to that player anonymously as well.
 We define the notion of anonymous entanglement, which may be useful in other scenarios as well:
\begin{definition}
If two anonymous players A and B share entanglement, we speak of \emph{anonymous entanglement (AE)}.
\end{definition}
\begin{definition}
If two players A and B share entanglement, where one of them is anonymous,
we speak of \emph{one-sided anonymous entanglement (one-sided AE)}.
\end{definition}

It is possible to use shared entanglement together with classical
communication to send quantum information using
quantum teleportation~\cite{BBCJPW93}.
 Anonymous entanglement together with a protocol providing classical sender anonymity
thus forms a virtual channel between two
players who do not know who is sitting at the other end. This
allows for easy sender and receiver anonymity for the
transmission of qubits. Note that it is also possible to use anonymous
entanglement to obtain a \emph{secure} classical anonymous channel. Unlike ANON,
this provides security of the data as well.
Classically, such a virtual channel would have to be
emulated by exchanging a key anonymously.
We require that if all players are honest,
the sender and recipient succeed in establishing an EPR pair. Furthermore,
the protocol should achieve sender and receiver anonymity with regard
to the two parts of the shared state. If one or more players are dishonest,
they may disrupt the protocol.

\subsubsection{Protocol}
We use the same resource of shared states $\ket{\Psi}$ to
establish anonymous entanglement for transmitting information by
using an idea presented in the context of quantum
broadcast~\cite{hein:coin}. More general protocols are certainly
possible. For now, we assume that there are exactly two players,
sender $s$ (Alice) and receiver $r$ (Bob), among the $n$ players
interested in sharing an EPR pair. If more players are interested,
they can use a form of collision detection described later.

\begin{protocol}{AE}{Shared state $(\ket{0^{n}} + \ket{1^{n}})/\sqrt{2}$.}\index{AE protocol}
\item Alice ($s$) and Bob ($r$) don't do anything to their part of the state.
\item Every player $j \in V \backslash \{s,r\}$
\begin{list}{-}{}
\item Applies a Hadamard transform to his qubit.
\item Measures this qubit in the computational basis with outcome $m_j$.
\item Broadcasts $m_j$.
\end{list}
\item $s$ picks a random bit $b \in_R \01$ and broadcasts $b$.
\item $s$ applies a phase flip $\sigma_z$ to her qubit if $b = 1$.
\item $r$ picks a random bit $b' \in_R \01$ and broadcasts $b'$.
\item $r$ applies a phase flip $\sigma_z$ to his qubit, if $b \oplus \bigoplus_{j \in V \backslash \{s,r\}} m_j = 1$.
\end{protocol}

\subsubsection{Correctness}
The shared state after the $n-2$ remaining players applied the
Hadamard transform becomes:
\begin{eqnarray*}
&&I_A \otimes I_B \otimes H^{\otimes (n-2)} \left(\frac{1}{\sqrt{2}}
(\ket{0^n} + \ket{1^n})\right) =\\
&=&\frac{1}{\sqrt{2^{n-1}}}\sum_{x
\in \01^{n-2}} (\ket{00}\ket{x} + (-1)^{|x|} \ket{11}\ket{x}).
\end{eqnarray*}
All players except Alice and Bob measure this state. The state for
them is thus $(\ket{00} + (-1)^{|x|} \ket{11})/\sqrt{2}$.
After Alice's phase flip the system is in state 
$(\ket{00} + (-1)^{|x| \oplus b} \ket{11})/\sqrt{2}$. The sum
of the measurements results gives $|x| = \bigoplus_{j \in V \backslash \{s,r\}} m_j$.
Thus Bob can correct the state to 
$(\ket{00} + \ket{11})/\sqrt{2}$ as desired.

\subsubsection{Anonymity}
The measurement outcomes are random. Thus, the players obtain no information during the measurement step.
Likewise, the bits broadcast by Alice and Bob are random. Thus both of them remain hidden.
Note that the protocol is resistant to collusions of up to $n-2$
players: The combined measurement outcomes still do not carry any
information about Alice and Bob.

\subsection{Sending Qubits}\label{quantum}

Let's return to the dinner table once more. After they have been
dining for hours on end, Bob, the waiter, finally shows up and
demands that the bill is paid. Alice, one of the dinner guests, is
indeed willing to pay using her novel quantum coins, however,
does not want to reveal this to her colleagues. The goal is now to transmit
an arbitrary qubit and not mere classical information. 
As before, we ask that our protocol achieves sender and receiver anonymity
and is traceless. Furthermore, if all players are honest, the receiver
should obtain the qubit sent. Note that unlike in the classical case,
we do not require that all honest players hold the same qubit at the end of the
protocol. This would contradict the no-cloning property of quantum states.
Alice now uses the shared EPR pair to send a quantum coin
$\ket{\phi}$ to Bob via teleportation~\cite{nielsen&chuang:qc}.

\begin{protocol}{ANONQ($\ket{\phi}$)}{Shared states $(\ket{0^{n}} + \ket{1^{n}})/\sqrt{2}$}\index{ANONQ protocol}
\item The players run AE: Alice and Bob now share an EPR pair: $\ket{\Gamma} = (\ket{00} + \ket{11})/\sqrt{2}$
\item Alice uses the quantum teleportation circuit with input $\ket{\phi}$ and EPR pair
$\ket{\Gamma}$, and obtains measurement outcomes $m_0,m_1$.
\item The players run ANON($m_0$) and ANON($m_1$) with Alice being the sender.
\item Bob applies the transformation described by $m_0, m_1$ on his part 
of $\ket{\Gamma}$ and obtains $\ket{\phi}$.
\end{protocol}

If all players are honest, after step 1, Alice and Bob share the
state $\ket{\Gamma} = (\ket{00} + \ket{11})/\sqrt{2}$ anonymously.
The correctness condition is thus satisfied by the correctness of
quantum teleportation. As discussed earlier, AE and ANON($b$) do
not leak any information about Alice or Bob. Since no additional
information is revealed during the teleportation step, it follows
that ANONQ($\ket{\phi}$) does not leak any information either and
our anonymity condition is satisfied. In our example, we only wanted
Alice to perform her payment anonymously, whereas Bob is known to all players. 
Our protocol also works, however, if Alice does not know the identity of Bob.

\section{Dealing with multiple senders}\label{multiple senders}

So far, we have assumed that only a single person is sending in any one round. In reality, many
users may wish to send simultaneously, leading to collisions.\index{collisions}
A user can easily detect a collision if it changes the classical outcome of the
transmission.
Depending on the application this may be sufficient. However, it may
be desirable to detect collisions leading to the same outcome.
This is important if we want to know the value of each of the bits sent and not
only their overall parity.

The simplest way to deal with collisions is for the user to wait a random number of rounds,
before attempting to resend the bit. This method was suggested by Chaum~\cite{chaum:dc}\index{Chaum, D.} and is generally known
as ALOHA~\cite{tanenbaum:cn}. Unfortunately this approach is rather wasteful, if many players try
to send simultaneously. Alternatively one could use a reservation map technique based on collision detection
similar to what was suggested by Pfitzmann et al.~\cite{pfitzmann:thesis}: For 
this one uses $n$ applications of collision detection (of $\ceil{\log n}+1$ rounds each) to 
reserve the following $n$ slots.

We will now present a simple quantum protocol to detect all kinds
of collisions, provided that no user tries to actively disrupt the
protocol. We use the same resource, namely shared entangled states
$\ket{\Psi}$. The important point of this protocol is that it
is traceless.

\subsection{Protocol}
Before each round of communication, the $n$ players run a $(\ceil{\log n} + 1)$-round test
to check, whether a collision would occur. For this they require $\ceil{\log n} + 1$ additional
states of the form $\ket{\Psi} = (\ket{0^n} + \ket{1^n})/\sqrt{2}$. Each  state is rotated before the
start of the collision detection protocol. Let
$$
U_j = R_z(-\pi/2^j) \otimes I^{\otimes (n-1)} = e^{i
\frac{\pi}{2^{j+1}}} \left(\begin{array}{cc} 1 & 0 \\ 0 &
e^{-i\pi/2^j}\end{array}\right) \otimes I^{\otimes (n-1)}
$$
and map the $j$th state to $\ket{t_j} = U_j\ket{\Psi}$.
This could for example be done by a dedicated
player or be determined upon distribution of the entangled states $\ket{\Psi}$.

\begin{protocol}{Collision Detection}{$\ceil{\log n} + 1$ states $\ket{\Psi} = (\ket{0^n} + \ket{1^n})/\sqrt{2}$}
\item A designated player prepares $\ceil{\log n} + 1$ states by rotations:\newline
      For $0 \leq j \leq \ceil{\log n}$, he applies $R_z(-\pi/2^j)$ to his
      part of one $\ket{\Psi}$ to create $\ket{t_j}$.
\item In round $0 \leq j \leq \ceil{\log n}$ each of the $n$ players
\begin{list}{-}{}
\item Applies $R_z(\pi/2^j)$ to his part of the state $\ket{t_j}$, if he wants to send.
\item Applies a Hadamard transform to his part of the state.
\item Measures in the computational basis.
\item Announces his measurement result to all other players.
\item Counts the total number of 1's, $k_j$, in the measurement results.
\item If $k_j$ is odd, concludes a collision has occurred and the protocol ends.
\end{list}
\item If all $k_j$ are even, exactly 1 player wants to send.
\end{protocol}

\subsection{Correctness and Privacy}
Let's first take an informal look, why this works.
In round $j$ with $0 \leq j \leq \ceil{\log n}$, each user who wishes to send applies a rotation
described by $R_z(\pi/2^j)$ to his part of the state. Note that if exactly one user tries
to send, this simply rotates the global state back to the original state
$\ket{\Psi} = (\ket{0^n} + \ket{1^n})/\sqrt{2}$. If $k > 1$ users try to send,
we can detect the collision in round $j$ such that $k = 2^j m + 1$ where $m \in \Natural$ is odd:
First $\ket{t_j}$ is rotated back to $\ket{\Psi}$ by the first of the $k$ senders.
The state is then rotated further by an angle of $(\pi/2^j) \cdot 2^j m = m \pi$.
But
$$
R_z(m\pi) = e^{-i \frac{m\pi}{2}}
\left(\begin{array}{cc}
1 & 0 \\
0 & e^{i m\pi}
\end{array}\right) =
\pm i
\left(\begin{array}{cc}
1 & 0 \\
0 & -1
\end{array}\right)
$$ applied to $\ket{\Psi}$ gives $\ket{\Psi'} = \pm i(\ket{0^n} - \ket{1^n})/\sqrt{2}$,
where we can ignore the global phase. The users now all apply a Hadamard transform
to their part of the state again, measure and broadcast their measurement results
to all players. As before, they can distinguish between $\ket{\Psi}$ and $\ket{\Psi'}$,
by counting the number of 1's in the outcome. If the number of users who want to
send in round $j$ is not of the form $2^jm + 1$,
the players may observe an even or odd number of 1's.
The crucial observation is that in $\ceil{\log n} + 1$ rounds, the players will obtain
$\ket{\Psi'}$ at least once, if more than one user wants to send, which they can detect.
If no phase flip has been observed in all rounds of the collision
detection protocol, the players can be sure there is exactly one sender.
The key to this part of the protocol is the following simple observation:

\begin{lemma}\label{form}
For any integer $2 \leq k \leq n$, there exist unique integers $m$
and $j$, with $m$ odd and $0 \leq j \leq \ceil{\log n}$, such that
$k = 2^j m + 1$.
\end{lemma}

\begin{proof}
By the fundamental theorem of arithmetic we can write $k - 1 = 2^j
m$ for unique $j,m \in \Natural$ where $m$ is odd. We have $j \leq
\ceil{\log n}$, since $2 \leq k \leq n$. Thus $k = 2^j m + 1$.
\end{proof}

\begin{corollary}
$\ceil{\log n}+1$ rounds, using one state $(\ket{0^n} +
\ket{1^n})/\sqrt{2}$ each, are sufficient to detect $2 \leq k \leq
n$ senders within a group of $n$ players.
\end{corollary}

\begin{proof}
Using Lemma~\ref{form} we can write $k = 2^j m + 1$ with $0 \leq j
\leq \ceil{\log n}$. In round $j$ the final state will be
$R_z((2^j m) \cdot (\pi/2^j))\ket{\Psi} = R_z(m\pi)\ket{\Psi} =
\pm i(\ket{0^n} - \ket{1^n})/\sqrt{2}$, which the players can
detect.
\end{proof}
\noindent
There exists a classical
protocol already suggested by Pfitzmann et al.~\cite{pfitzmann:birgit} using $O(n^2 \log n)$ bits
of private shared randomness. However, this protocol is not traceless as desired by our protocol.
Our protocol preserves anonymity and is traceless by the same argument used
in Section~\ref{anon security}.

When sending quantum states, collisions are not so easy to detect, since
they do not change the outcome noticeably.
The protocol to establish anonymous entanglement relies on the fact that 
only two players refrain from measuring. We thus require some coordination 
between the two players. Here, we can make use of the same collision 
detection protocol as we used to send classical bits:
First run the collision detection protocol to determine the sender.
The sender again expresses his interest in
indicating that he wants to send by employing rotations.
Then perform another application of collision detection
for the receiver.

\section{Conclusions and Future Work}

We have presented a protocol for achieving anonymous transmissions using shared quantum states together with
a classical broadcast channel. The main feature
of this protocol is that, unlike all classical protocols, it prevents later reconstruction of the sender.
This indicates that shared entangled states are very well suited to achieve anonymity. Perhaps similar
techniques could also play an important role in other protocols where such a traceless property is
desirable.

Our protocol is a first attempt at providing anonymous transmissions with this particular property.
More efficient protocols may be possible. Perhaps a different form of quantum resource gives
an additional advantage.
However, we believe that our protocol is close to optimal for the given resources.
We have also not considered the possibility of allowing quantum communication
between the players, which could be required by more efficient protocols.
It is also open whether a better form of collision detection and protection against malicious disruptors is
possible. The states used for our collision detection protocol are hard 
to prepare if $n$ 
is very large. Furthermore, using shared entangled states, it is always possible for a malicious user to measure his qubit in
the computational basis to make further transmissions impossible.

So far, we
have simply assumed that the players share a certain quantum
resource. In reality, however, this resource would need to be
established before it can be used. This would require quantum
communication among the players in order to distribute the
necessary states and at least classical communication for
verification purposes. The original DC-net protocol suffers from a
similar problem with regard to the distribution of shared
keys, which
is impossible to do from scratch using only classical
channels~\cite{nielsen&chuang:qc}. Some quantum states on the other hand have
the interesting property that the players can create and test
the states among themselves, instead of relying on a trusted third
party.

\section{Acknowledgments}
We thank Andreas Pfitzmann for sending us a copy
of~\cite{pfitzmann:thesis} and~\cite{pfitzmann:isdn}. We also
thank Ronald de Wolf, Louis Salvail and Renato Renner for useful comments. SW
thanks the CQC Cambridge for their hospitality.

\end{document}